\documentclass[12pt,a4paper]{article}
\pdfoutput=1
\usepackage[utf8]{inputenc}
\usepackage{epsf,amsmath,amssymb,graphicx,dcolumn}
\usepackage{caption}
\usepackage{subcaption}
\usepackage{scalefnt,ulem,pstricks}
\usepackage{booktabs,multirow,tabularx}
\usepackage{cite,hyperref}
\usepackage{cleveref}
\usepackage{color}
\usepackage{rotating}
\usepackage{cancel}
\usepackage{microtype}
\usepackage[titletoc,title]{appendix}
\usepackage[numbers,sort&compress]{natbib}

\providecommand{\href}[2]{#2}

\def\ltap{\raisebox{-.6ex}{\rlap{$\,\sim\,$}} \raisebox{.4ex}{$\,<\,$}} 
\def\gtap{\raisebox{-.6ex}{\rlap{$\,\sim\,$}} \raisebox{.4ex}{$\,>\,$}}

\newcommand\as{\alpha_{\mathrm{S}}}

\def\to{\rightarrow}

\def\mz{m_Z}

\newcommand{\TeV}{\ensuremath{\rm TeV}}

\newcommand\Matrix{{\sc Matrix}}
\newcommand\aclr{{\sc aclr}}
\newcommand\Munich{{\sc Munich}}
\newcommand\OpenLoops{{\sc OpenLoops}}
\newcommand\Recola{{\sc Recola}}
\newcommand\Collier{{\sc Collier}}

\newcommand{\CutTools}{{\sc CutTools}}
\newcommand{\OneLOop}{{\sc OneLOop}}
\newcommand{\qt}{\ensuremath{q_T}}

\newcommand{\fig}[1]{Figure~\ref{#1}}

\newcommand{\tab}[1]{Table~\ref{#1}}

\newcommand{\sct}[1]{Section~\ref{#1}}

\def\refta#1{\mbox{Table~\ref{#1}}}

\def\refse#1{\mbox{Section~\ref{#1}}}

\newcommand{\citere}[1]{\mbox{Ref.~\cite{#1}}}
\newcommand{\citeres}[1]{\mbox{Refs.~\cite{#1}}}

\newcommand{\rcut}{\ensuremath{r_{\mathrm{cut}}}}
\newcommand{\zz}{\ensuremath{ZZ}}
\newcommand{\ww}{\ensuremath{W^+W^-}}

\newcommand{\z}{\ensuremath{Z}}

\renewcommand{\gg}{\ensuremath{gg}} 
\newcommand{\qqx}{\ensuremath{q\bar{q}}}
\newcommand{\qg}{\ensuremath{qg}}

\newcommand{\abbrev}{}
\newcommand{\nnlo}{\text{\abbrev NNLO}}

\newcommand{\qcd}{{\abbrev QCD}}

\newcommand{\LO}{\ensuremath{{\rm{\abbrev LO}}}}
\newcommand{\lo}{\LO}
\newcommand{\NLO}{\ensuremath{{\rm{\abbrev NLO}}}}
\newcommand{\nlo}{\NLO}
\newcommand{\qqNNLO}{\ensuremath{q\bar{q}{\rm{\abbrev NNLO}}}}
\newcommand{\ggLO}{\ensuremath{gg{\rm{\abbrev LO}}}}
\newcommand{\ggNLO}{\ensuremath{gg{\rm{\abbrev NLO}}}}
\newcommand{\ggNLOgg}{\ensuremath{gg{\rm{\abbrev NLO}}_{gg}}}
\newcommand{\NNLO}{\ensuremath{{\rm{\abbrev NNLO}}}}
\newcommand{\nNNLO}{\ensuremath{{\rm n{\abbrev NNLO}}}}

\newcommand{\dyZZ}{\ensuremath{\Delta y_{ZZ}}}

\newcommand{\mll}{\ensuremath{m_{\ell^+\ell^-}}}

\newcommand{\mfourl}{\ensuremath{m_{4\ell}}}

\newcommand{\ptzone}{\ensuremath{p_{T,Z_1}}}
\newcommand{\ptztwo}{\ensuremath{p_{T,Z_2}}}
\newcommand{\ptzi}{\ensuremath{p_{T,Z_i}}}

\newcommand{\ptlone}{\ensuremath{p_{T,\ell_1}}}

\newcommand{\dRll}{\ensuremath{\Delta R_{\ell\ell}}}

\usepackage{amssymb}

\usepackage{bm, amsbsy}

\begin{document} 
\hypersetup{pageanchor=false}
\begin{titlepage}
\renewcommand{\thefootnote}{\fnsymbol{footnote}}
\begin{flushright}
  CERN-TH-2018-240\\
     ZU-TH 42/18
     \end{flushright}
\par \vspace{10mm}

\begin{center}
{\Large \bf $\boldsymbol{\zz}$ production at the LHC: NLO QCD corrections\\[0.2cm] to the loop-induced gluon fusion channel}
\end{center}

\par \vspace{2mm}
\begin{center}
  {\bf Massimiliano Grazzini}$^{(a)}$, {\bf Stefan Kallweit}$^{(b,c)}$,\\[0.2cm]
  {\bf Marius Wiesemann}$^{(b)}$ and {\bf Jeong Yeon Yook}$^{(a)}$

$^{(a)}$ Physik-Institut, Universit\"at Z\"urich, CH-8057 Z\"urich, Switzerland 

$^{(b)}$ TH Division, Physics Department, CERN, CH-1211 Geneva 23, Switzerland

$^{(c)}$ Universit\`{a} degli Studi di Milano-Bicocca, 20126, Milan, Italy

\end{center}

\begin{center} {\bf Abstract} \end{center}\vspace{-1cm}
\begin{quote}
\pretolerance 10000

We consider QCD radiative corrections to the production of four charged leptons in hadron collisions.
We present the computation of the next-to-leading order QCD corrections to the loop-induced gluon fusion contribution.
Our predictions include, for the first time, also the quark--gluon partonic channels.
The computed corrections, which are formally of ${\cal O}(\as^3)$, turn out to increase the loop-induced Born-level result
by an amount ranging from $75\%$ to $71\%$ as $\sqrt{s}$ ranges from $8$ to $13\,\TeV$.
We combine our result with state-of-the-art \NNLO{} corrections to the quark annihilation channel,
and present updated predictions for fiducial cross sections and distributions for this process.

\end{quote}

\vspace*{\fill}
\begin{flushleft}
November 2018

\end{flushleft}
\end{titlepage}
\hypersetup{pageanchor=true}

\section{Introduction}

The production of $Z$-boson pairs is one of the most relevant processes at the Large Hadron Collider (LHC).
Besides providing an important
test of the electroweak~(EW) sector of the Standard Model~(SM) at the \TeV{} scale, it
was instrumental for the discovery of the Higgs boson \cite{Aad:2012tfa,Chatrchyan:2012xdj}. As
the focus of Higgs physics moved from the discovery to the study of its properties,
$Z$-boson pair production played an essential role in the determination of the quantum numbers of the new
resonance~\cite{Aad:2013xqa,Chatrchyan:2013mxa}, in setting bounds on its width
(see e.g.\ \citeres{Khachatryan:2015mma,Aaboud:2018puo}),
and in constraining anomalous Higgs boson couplings \cite{Khachatryan:2016vau}.

At the leading order (LO) in the QCD coupling $\as$, $Z$-boson pairs are produced via quark annihilation.
Theoretical predictions for \zz{} production at next-to-leading order (\nlo{}) \qcd{} 
were obtained a long time ago for both on-shell \z{} bosons~\cite{Ohnemus:1990za,Mele:1990bq} and 
their fully leptonic final states~\cite{Ohnemus:1994ff,Campbell:1999ah,Dixon:1999di,Dixon:1998py}. 
Perturbative corrections beyond \NLO{} QCD are indispensable to reach the precision demanded by present \zz{} measurements. 
NLO EW corrections are known for stable \z{} bosons~\cite{Accomando:2004de,Bierweiler:2013dja,Baglio:2013toa} 
and including their leptonic decays with full off-shell effects~\cite{Biedermann:2016yvs,Biedermann:2016lvg}.
NLO QCD+EW results for the $2\ell 2\nu$ signature have been presented in \citere{Kallweit:2017khh},
and for the $\ell\ell\ell'\ell'$ signature with the inclusion of anomalous couplings in \citere{Chiesa:2018lcs}.
$\zz$+${\rm jet}$ production was computed at \NLO{} QCD~\cite{Binoth:2009wk}.
The loop-induced gluon fusion channel, which provides a separately finite ${\cal O}(\as^2)$ contribution and
is enhanced by the large gluon luminosity, has been known at LO for a long
time~\cite{Glover:1988rg,Dicus:1987dj,Matsuura:1991pj,Zecher:1994kb,Binoth:2008pr,Campbell:2011bn,Kauer:2013qba,Cascioli:2013gfa,Campbell:2013una,Ellis:2014yca,Kauer:2015dma}.
It was recently computed at \NLO{}~\cite{Caola:2015psa,Caola:2016trd,Alioli:2016xab} using the
two-loop helicity amplitudes for $\gg\to VV'$ of \citeres{Caola:2015ila,vonManteuffel:2015msa},
considering only the gluon--gluon~(\gg{}) partonic channel. 
\nnlo{} \qcd{} corrections to on-shell \zz{} production were first evaluated 
in \citere{Cascioli:2014yka}, and later in \citere{Heinrich:2017bvg}. Using the
two-loop helicity amplitudes for $\qqx\to VV'$~\cite{Gehrmann:2014bfa,Caola:2014iua,Gehrmann:2015ora},
fully differential \NNLO{} predictions in the
four-lepton channels ($\ell\ell\ell\ell$ and $\ell\ell\ell'\ell'$) were first presented 
in \citere{Grazzini:2015hta}, while in \citere{Kallweit:2018nyv} also the $2\ell2\nu$ signature was considered.

An analogous situation is the one of \ww{} production, for which \nnlo{} \qcd{}
corrections~\cite{Gehrmann:2014fva,Grazzini:2016ctr} to quark annihilation are available,
and \nlo{} \qcd{} corrections to the loop-induced gluon fusion contribution were computed recently~\cite{Caola:2015rqy}.
At present, experimental analyses for both $ZZ$ and $WW$ production treat the quark annihilation and loop-induced gluon fusion channels as if they were independent. As a result, data are compared to {\it ad hoc} combinations of NNLO calculations for the quark annihilation channel and NLO calculations for the loop-induced gluon fusion channel, often by using $K$-factors (see e.g.\ \citeres{Aaboud:2017rwm,Aaboud:2017qkn,Sirunyan:2017zjc,CMS:2016vww}).
However, it is well known that the quark--antiquark~($\qqx{}$) and \gg{} partonic channels mix through parton evolution,
and thus their independent treatment is not appropriate.  Moreover, already at \NNLO{} there are diagrams that
mix the two production mechanisms, thereby suggesting that a unified treatment would be desirable.
This is particularly important to consistently estimate the perturbative uncertainties
through variation of the renormalisation and factorisation scales.

In this paper we take a decisive step in this direction, by combining the \NNLO{} calculation in the quark annihilation
channel with the \NLO{} calculation of the loop-induced gluon fusion channel. For the first time, we also evaluate
the (anti)quark--gluon (\qg{}) contributions that enter the full \NLO{} corrections to the loop-induced channel.
We introduce an approximation of the full N$^3$LO corrections, denoted by ``\nNNLO{}'',
which represents the most advanced perturbative QCD prediction available at present
for this process. The new calculation will be available in an updated version of \Matrix{}~\cite{Grazzini:2017mhc}.

The paper is organised as follows. In \sct{sec:framework} we introduce our computational framework.
In \sct{sec:valid} we present a comparison of our results to those of \citere{Alioli:2016xab}.
In \sct{sec:results} we combine our computations of radiative corrections to the quark annihilation and
loop-induced gluon fusion channels, and present fiducial cross sections and distributions in $pp$ collisions
at $8$ and $13\,\TeV$. In \sct{sec:summary} we summarise our results.

\section{Calculation within the M{\small ATRIX} framework}
\label{sec:framework}

\begin{figure}[h]
\begin{center}
\begin{tabular}{ccccc}
\includegraphics[height=2.0cm]{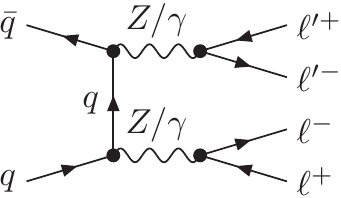}& \quad\quad\quad\quad &
\includegraphics[height=2.0cm]{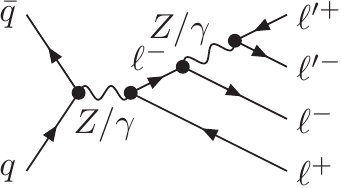}& \quad\quad\quad\quad &
\includegraphics[height=2.0cm]{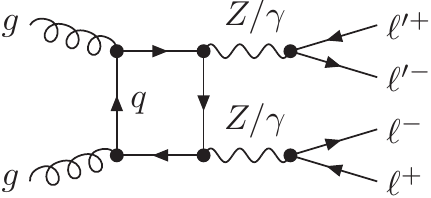} \\[0ex]
(a) & \quad & (b) & \quad & (c)
\end{tabular}
\end{center}
\caption[]{\label{fig:diag}{Sample Feynman diagrams for \zz{} production with four charged final-state leptons:
    tree-level diagrams of the quark annihilation channel in (a) and (b), loop-induced diagram of the gluon fusion channel in (c).}}
\end{figure}

We consider the four-lepton process
\begin{align}
pp \rightarrow \ell^+\ell^-\,\ell'^+\ell'^-+X,\nonumber
\end{align}
where, for simplicity, we assume the triggered lepton pairs to have different flavours ($\ell \neq \ell'$).
Representative
Born-level diagrams are shown in \fig{fig:diag}. Diagrams (a) and (b) are driven by quark annihilation and show double-resonant $t$-channel 
\zz{} production and single-resonant $s$-channel Drell--Yan topologies, respectively. Diagram (c) is instead driven by gluon fusion through a quark loop,
and it enters the calculation at \NNLO{} as it is of ${\cal O}(\as^2)$. However, this contribution is enhanced by the large gluon luminosity.
Up to \NLO{} the quark annihilation and loop-induced gluon fusion production processes do not mix. Until a few years ago,
the theoretical standard was to consider \NLO{}-accurate predictions for the quark annihilation channel, supplemented with the loop-induced gluon fusion contribution~\cite{Campbell:2011bn}.

\begin{figure}[t]
\begin{center}
\begin{tabular}{c}
\includegraphics[height=2.5cm]{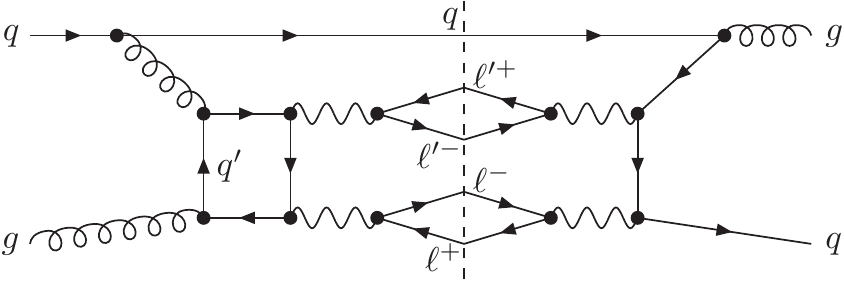}\\
\end{tabular}
\end{center}
\caption[]{\label{fig:mix}{Example of \NNLO{} interference between quark annihilation and loop-induced gluon fusion production mechanisms.}}
\end{figure}

\begin{figure}
\vspace{0.8cm}
\begin{center}
\begin{tabular}{ccc}
\includegraphics[height=2.5cm]{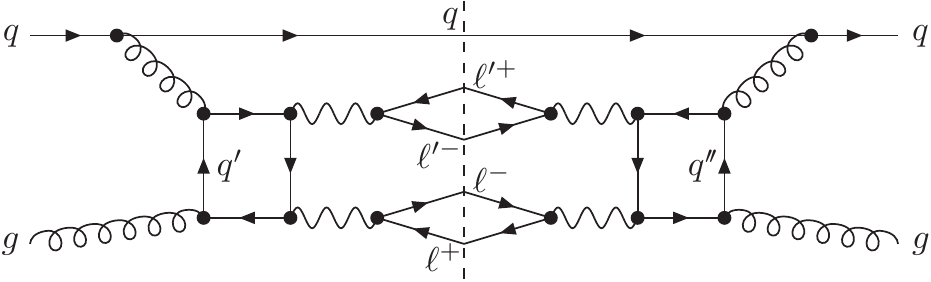}& \quad &
\includegraphics[height=2.5cm]{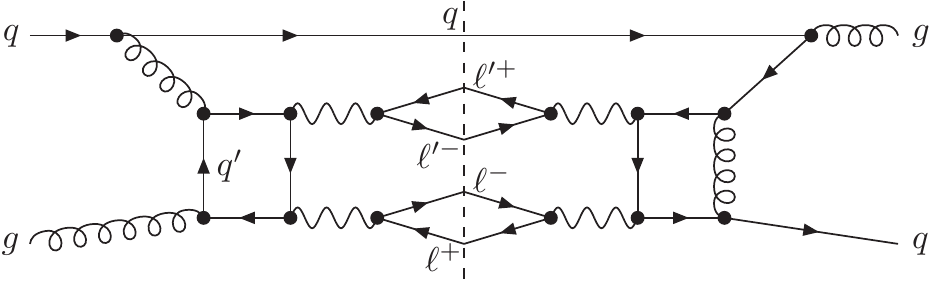}\\
(a) & \quad & (b)\\
\end{tabular}
\end{center}
\caption[]{\label{fig:qg}{Examples of N$^3$LO contributions in the \qg{} channel.}}
\end{figure}

Starting from \NNLO{}, the quark annihilation and loop-induced gluon fusion processes mix, and the distinction between the two production
mechanisms is questionable.
An example of an interference contribution is shown in \fig{fig:mix}.
A complete \NNLO{} computation of four-lepton production has been presented in \citeres{Grazzini:2015hta,Kallweit:2018nyv}.
At this order, the loop-induced gluon fusion contribution enters the cross section 
through the square of diagrams like the one in \fig{fig:diag}\,(c).
The fact that this ${\cal O}(\as^2)$ contribution is quite large and formally only \LO{} accurate
motivates the inclusion of \NLO{} corrections to the loop-induced gluon fusion channel, which are part of the N$^3$LO corrections.
We point out that the loop-induced contributions are not the only contributions to the $gg$ channel at N$^3$LO.
However, we expect the impact of the N$^3$LO non-loop-induced diagrams to be within the perturbative uncertainties estimated by studying scale variations at NNLO.
The same cannot be said for the loop-induced contributions.
The \NLO{} computation for a loop-induced process requires one-loop amplitudes with the emission of one additional
parton and two-loop contributions.
In \citeres{Caola:2015psa,Alioli:2016xab} the calculation has been performed by considering
only the \gg{} partonic channel.
Here we extend the above calculation by including also the \qg{} initiated contributions.\footnote{We note that there are also \qqx{} initiated contributions to the 
loop-induced production mechanism at $\mathcal{O}(\as^3)$, which are separately finite. We found them to be completely negligible and ignore them in the following. 
Our results include all numerically relevant partonic channels of the NLO corrections to the loop-induced gluon fusion contribution.}
We note that at N$^3$LO we only include diagrams
with closed fermion loops (see \fig{fig:qg}\,(a));
all other contributions that would enter a complete N$^3$LO calculation (see \fig{fig:qg}\,(b) for example) cannot be consistently accounted for at present.
Our approximation includes all contributions at $\mathcal{O}(\as^2)$
together with the complete \NLO{} corrections to the loop-induced gluon fusion channel at $\mathcal{O}(\as^3)$.
As such, besides providing the maximum perturbative information available at present for this process,
our calculation can be used to obtain a consistent estimate of perturbative uncertainties
through the customary procedure of studying scale variations.

Our calculation is carried out within the computational framework \Matrix{}~\cite{Grazzini:2017mhc}. \Matrix{} features
a fully general implementation of the \qt{}-subtraction formalism~\cite{Catani:2007vq} and allowed us to compute
\NNLO{} QCD corrections to a large number of colour-singlet processes at hadron
colliders~\cite{Grazzini:2013bna,Grazzini:2015nwa,Cascioli:2014yka,Grazzini:2015hta,Gehrmann:2014fva,Grazzini:2016ctr,Grazzini:2016swo,Grazzini:2017ckn,deFlorian:2016uhr,Grazzini:2018bsd}.\footnote{It was
also used in the NNLL+NNLO computation of \citere{Grazzini:2015wpa}, and in the NNLOPS computation of \citere{Re:2018vac}.}
The core of the \Matrix{} framework is the Monte Carlo program \Munich{}, which is capable of
computing both \NLO{} QCD and \NLO{} EW~\cite{Kallweit:2014xda,Kallweit:2015dum} corrections
to arbitrary SM processes~\cite{munich}.

As in previous \Matrix{} calculations, in our computation of the \NLO{} corrections to the $\gg\to 4\ell$ process, all the required one-loop amplitudes are evaluated with 
\OpenLoops{}\footnote{\OpenLoops{} relies on the fast and stable
tensor reduction of \Collier{}~\cite{Denner:2014gla,Denner:2016kdg},
supported by a rescue system based on quad-precision
\CutTools~\cite{Ossola:2007ax} with \OneLOop~\cite{vanHameren:2010cp}
to deal with exceptional phase-space
points.
All relevant loop-induced amplitudes with correlators will be available in an upcoming publication of 
\OpenLoops{}2~\cite{openloops2}.}~\cite{Cascioli:2011va,Buccioni:2017yxi}.
To the purpose of validating our results for the loop-induced contribution, we have used also the independent matrix-element generator \Recola{}~\cite{Actis:2016mpe,Denner:2017wsf}, finding complete agreement.

At two-loop level, we use the $\gg\to VV'$ helicity amplitudes of \citere{vonManteuffel:2015msa},
and implement the corresponding four-lepton final states, accounting for spin correlations
and off-shell effects.
The \NLO{} calculation is performed by using the 
Catani--Seymour dipole-subtraction method~\cite{Catani:1996jh,Catani:1996vz} and 
also with $q_T$ subtraction~\cite{Catani:2007vq}, which provides an additional cross-check of our results.
  
\section{Validation}
\label{sec:valid}

The \NLO{} corrections to the loop-induced gluon fusion contribution have been first computed
in \citeres{Caola:2015psa,Alioli:2016xab}, where the \qg{} partonic channels were neglected.
The results of \citere{Caola:2015psa} are provided with only two significant digits, and without any information
on numerical or systematic uncertainties. Although we were able to reproduce
their results at this level of precision, they are not particularly suitable for a detailed comparison.
More precise results are stated in \citere{Alioli:2016xab}, and with the statistical errors provided
in private communication a reasonable technical comparison could be performed.
The calculations in \citeres{Caola:2015psa,Alioli:2016xab} are carried out by using five massless quark flavours ($N_f$),
and the contributions of both top-quark loops and triangle diagrams are omitted.\footnote{The fermionic triangle diagrams
vanish in the sum over a massless quark generation. Correspondingly, neglecting the remaining bottom-quark
contribution is a reasonable ansatz if the top-quark contribution cannot be taken into account.}
Furthermore, the \qg{} initiated subprocesses were neglected.
The loop-induced gluon fusion contribution (denoted by \ggLO{} in the following) and its \NLO{} corrections
restricted to the \gg{} partonic channel (denoted by \ggNLOgg{} in the following) are both computed with $N_f=5$
NNPDF3.0~\cite{Ball:2014uwa} \NLO{} parton distribution functions with $\as(m_Z)=0.118$.

For validation, we have repeated the calculation of \citere{Alioli:2016xab},
using exactly the same setup:
Cuts are only applied on the invariant mass of the opposite-sign same-flavour (OSSF) lepton pairs (\mll{}) and on
the four-lepton invariant mass (\mfourl{}), which are required to fulfil
\begin{align}
5\,{\rm GeV} < \mll{} < 180\,{\rm GeV}\,,\qquad
60\,{\rm GeV} < \mfourl{} < 360\,{\rm GeV}\, .
\end{align}
The comparison of our results to those in Table I of \citere{Alioli:2016xab} (denoted by \aclr{})
is reported in \tab{table:ggcomp}.
Both settings of \citere{Alioli:2016xab} for the renormalisation ($\mu_R$) and factorisation ($\mu_R$) scales are considered: fixed values 
$\mu_R=\mu_F=\mu_0\equiv m_Z$, with $m_Z$ being the $Z$-boson mass, and dynamical values $\mu_R=\mu_F=\mu_0\equiv \mfourl/2$. 
The quoted uncertainties are estimated from customary 7-point 
scale variations by a factor of two, with the constraint $0.5\le \mu_R/\mu_F\le 2$.
For both our and the \aclr{} results we provide statistical errors on the respective last digit(s) inside parentheses.

\renewcommand{\baselinestretch}{1.5}
\begin{table}[h]
\begin{center}
\small
\begin{tabular}{|c|c| c | c || c | c |}
    \hline
 & & \multicolumn{2}{c||}{$\mu=m_{4l}/2$} & \multicolumn{2}{c|}{$\mu=\mz$\phantom{\Big|}}    \\
  \hline
  $\sqrt{s}$
 &$~$
 &ggLO
 &ggNLO$_{gg}$
 &ggLO
 &ggNLO$_{gg}$\phantom{\Big|} \\
  \hline
  \phantom{1}8 TeV\phantom{\Big|}
  & \aclr{}
  & $1.6018(5)^{+0.41}_{-0.30}$
  & $2.980(5)^{+0.51}_{-0.41}$
  & $1.6181(6)^{+0.42}_{-0.31}$
  & $2.978(4)^{+0.49^4}_{-0.40}$
  \\
  & \Matrix{}
  & {$1.6023(4)^{+0.41}_{-0.30}$}
  & {$2.987(3)^{+0.51}_{-0.42}$}
  & {$1.6188(3)^{+0.42}_{-0.31}$}
  & {$2.985(3)^{{+0.49\phantom{^4}}}_{-0.40}$}
  \\
  \hline
  13 TeV          \phantom{\Big|}
  & \aclr{}
  & $3.8467(13)^{+0.97}_{-0.70}$
  & $6.984(8)^{+1.14}_{-0.94}$
  & $3.9429(14)^{+0.98}_{-0.71}$
  & $7.22(18)^{+1.04}_{-1.04}$
  \\
  & \Matrix{}
  & {$3.8486(8)^{+0.97}_{-0.70}\phantom{0}$}
  & {$7.016(7)^{+1.15}_{-0.95}$}
  & {$3.9454(8)^{+0.98}_{-0.71}\phantom{0}$}
  & {$7.068(7)^{+1.11}_{-0.93}$}
  \\
  \hline
\end{tabular}
\end{center}
\caption{\label{table:ggcomp}Comparison of our \Matrix{} results with those of \citere{Alioli:2016xab}.\protect\footnotemark}
\end{table}

\renewcommand{\baselinestretch}{1.0}

Taking into account the numerical errors of the \aclr{} results,
which are typically of comparable size as our errors, we find
that all $8\,\TeV$ results are in perfect statistical agreement, corresponding to discrepancies of only a few per mille of the
respective \ggNLOgg{} predictions.
The agreement between the \ggNLOgg{} results at $13\,\TeV$ turns out to be slightly worse:
While the exceptionally large discrepancy of $-2.1\%$ with $\mu_0=m_Z$ as central scale choice is fully covered by the
statistical \aclr{} error within one standard deviation,
the few-permille discrepancy in the result for $\mu_0= \mfourl/2$ corresponds to a discrepancy of three standard deviations.
We note that the calculation of \citere{Alioli:2016xab} uses a technical cut $p_T^{ZZ}>0.5$ GeV on the transverse momentum
of the four-lepton system, in order to avoid instabilities in the one-loop matrix elements.
Also our calculation involves technical cuts based on parameters controlling the minimum invariant masses of parton pairs and
the internal stability estimate of the \OpenLoops{} amplitudes.
By varying the above technical parameters we estimate the systematic uncertainties affecting our \NLO{} results
in \refta{table:ggcomp} to be at the few-permille level. Indeed, this level of precision is confirmed by our alternative
implementation using \qt{} subtraction.
This additional source of uncertainties in the two predictions could explain the slight discrepancy
we observe in the \NLO{} result at $13\,\TeV$ with $\mu_0=\mfourl/2$, which is poorly covered by the statistical errors only.
We stress that for the purpose of phenomenological applications all the observed differences are subleading since a percent
effect in the loop-induced gluon fusion channel only leads to a permille effect on the complete inclusive and fiducial four-lepton
cross sections due to the dominance of the quark annihilation channel.
We also note that we were able to qualitatively (on the level of the plots) reproduce all differential distributions
shown in \citeres{Caola:2015psa,Alioli:2016xab}.
\footnotetext{A typo in the upper scale variation of the \ggNLOgg{} result for $\sqrt{s}=8\,\TeV$, $\mu_0= \mfourl/2$ of \citere{Alioli:2016xab} has been corrected.}

\section{Results}
\label{sec:results}

\subsection{Setup}

We present predictions for $pp\rightarrow e^+e^-\mu^+\mu^-$ production at $8$ and $13\,\TeV$.
For the EW parameters we employ the $G_\mu$ scheme and
set $\alpha=\sqrt{2}\,G_\mu m_W^2(1-m_W^2/m_Z^2)/\pi$. Contrary to \sct{sec:valid}, we compute the EW mixing angle as
$\cos\theta_W^2=(m_W^2-i\Gamma_W\,m_W)/(m_Z^2-i\Gamma_Z\,m_Z)$
and use the complex-mass scheme~\cite{Denner:2005fg} throughout.
The EW inputs are set to the PDG~\cite{Patrignani:2016xqp} values: $G_F =
1.16639\times 10^{-5}$\,GeV$^{-2}$, $m_W=80.385$\,GeV,
$\Gamma_W=2.0854$\,GeV, $m_Z = 91.1876$\,GeV, $\Gamma_Z=2.4952$\,GeV,
$m_H = 125$\,GeV, and $\Gamma_H = 0.00407$.
The on-shell top-quark mass is set to $m_t = 173.2$\,GeV, and $\Gamma_t=1.44262$ is used.
Except for virtual two-loop contributions, the full dependence on massive top quarks 
is taken into account everywhere in the computation.
For the $\qqx\to ZZ$ subprocess, the contribution of top quarks in the two-loop corrections is not known.
Since the quantitative impact of the two-loop diagrams with a light fermion loop is extremely small \cite{Cascioli:2014yka}, the two-loop diagrams involving
a top-quark can be safely neglected.
Disregarding Higgs boson exchange\footnote{Quantitative comments on the impact of Higgs boson diagrams are postponed to \refse{sec:fid}.},
top-quark loops contribute about $2\%$ to the \ggLO{} cross section and should thus be included.
However, also for the $\gg\to ZZ$ subprocess, the top-quark contribution in the two-loop amplitude is unknown.
Here we approximate top-quark effects through a reweighting of the massless two-loop result by the \LO{} (one-loop) amplitude with full top-quark mass dependence.
Diagrams involving the Higgs boson are consistently included at each perturbative order, except for the two-loop contributions, where we employ the same approximation as for the massive top loops.

For each perturbative order we use the corresponding set of $N_f=5$ 
NNPDF3.0~\cite{Ball:2014uwa} parton distribution functions~(PDFs)
with $\as(m_Z)=0.118$. The loop-induced gluon fusion contribution and its \NLO{} corrections are always computed with \NNLO{} PDFs.
The renormalisation and factorisation scales are set to
half of the invariant mass of the four-lepton system, $\mu_R=\mu_F=\mu_0\equiv\mfourl/2$.
Residual uncertainties are estimated from customary 7-point 
scale variations by a factor of two, with the constraint $0.5\le \mu_R/\mu_F\le 2$.

We use the selection cuts adopted by the ATLAS collaboration at $8\,\TeV$ in \citere{Aaboud:2016urj},
which are summarized in \refta{tab:cuts}.\footnote{For simplicity, we employ the same setup at $13$\,TeV.}
The fiducial cuts involve standard requirements 
on the transverse momenta and pseudo-rapidities of the leptons, a pair-wise separation 
in $\Delta R=\sqrt{\Delta \eta^2+\Delta \phi^2}$ between all possible leptons (independently of their flavours and charges),
and a window in the invariant mass of reconstructed $Z$ bosons 
around the $Z$-pole.

\renewcommand{\baselinestretch}{1.5}
\begin{table}[!h]
\begin{center}
\begin{tabular}{c}
\toprule
definition of the fiducial volume for $pp\to e^+e^-\mu^+\mu^-+X$\\
\midrule
$p_{T,e/\mu}>7$\,GeV,  \quad one electron with $|\eta_e|<4.9$, \quad the others $|\eta_e|<2.5$,  \quad$|\eta_\mu|<2.7$\\
$\Delta R_{ee/\mu\mu} >0.2$, \quad$\Delta R_{e\mu} >0.2$, \quad$66\,\textrm{GeV}\le m_{e^+e^-/\mu^+\mu^-} \le 116$\,GeV,\\
\bottomrule
\end{tabular}
\end{center}
\renewcommand{\baselinestretch}{1.0}
\caption{\label{tab:cuts} 
Phase-space definitions of the \zz{} measurements by ATLAS at 8\,TeV~\cite{Aaboud:2016urj}.}
\vspace{-0.5cm}
\end{table}

\renewcommand{\baselinestretch}{1.0}

\subsection{Fiducial cross section and distributions}
\label{sec:fid}

We briefly introduce the notation used throughout this section:
The loop-induced gluon fusion channel contributes at ${\cal O}(\as^2)$, and is denoted by \ggLO{} in the following. 
The \NNLO{} result for the quark annihilation channel, i.e.\ without the loop-induced contribution, is referred to as \qqNNLO{}.
The complete loop-induced contribution at \NLO{} is labelled \ggNLO{}, while
its restriction to the \gg{} partonic channel is dubbed as \ggNLOgg{}, i.e.\ the difference between these two predictions corresponds to the newly computed contribution from the \qg{} channels.
As discussed in the Introduction,
the \NLO{} corrections to the loop-induced contribution
are only a part of the complete  N$^3$LO computation.
However, these corrections are sizeable, and the loop-induced gluon fusion production mechanism is known to be only poorly described at ${\cal O}(\as^2)$, namely at its effective \LO{}. It is thus reasonable
to construct an approximation of the complete N$^3$LO cross section based on the inclusion of only these ${\cal O}(\as^3)$ corrections.
This approximation is denoted by \nNNLO{}.

\renewcommand{\baselinestretch}{1.5}
\begin{table}[t]
\begin{minipage}{\textwidth}  
\begin{center}
\begin{tabular}{|l|rcr|rcr|}
\hline
\multicolumn{1}{|c|}{$\sqrt{s}$} & 8\,TeV  & & 13\,TeV & 8\,TeV & & 13\,TeV  \\
\hline
& \multicolumn{3}{c|}{$\sigma$\,[fb]} &\multicolumn{3}{c|}{$\sigma/\sigma_{\rm NLO}-1$} \\
\hline
\lo{}      
& $8.1881(8)_{-3.2\%}^{+2.4\%}$ 
&
& $13.933(1)_{-6.4\%}^{+5.5\%}$
& $-27.5\%$ 
&
& $-29.8\%$ \\
\nlo{} 
& $11.2958(4)_{-2.0\%}^{+2.5\%}$
&
& $19.8454(7)_{-2.1\%}^{+2.5\%}$
& $0\%$ 
&
& $0\%$ \\
\qqNNLO{}
& $12.09(2)_{-1.1\%}^{+1.1\%}$
&
& $21.54(2)_{-1.2\%}^{+1.1\%}$
& $+7.0\%$ 
&
& $+8.6\%$\\
\hline
& \multicolumn{3}{c|}{$\sigma$\,[fb]} &\multicolumn{3}{c|}{$\sigma/\sigma_{\rm {ggLO}}-1$} \\
\hline
\ggLO{}
& $0.79355(6)_{-20.9\%}^{+28.2\%}$
&
& $2.0052(1)_{-17.9\%}^{+23.5\%}$
& $0\%$ 
&
& $0\%$ \\ 
\ggNLOgg{}      
& $1.4787(4)_{-13.1\%}^{+15.9\%}$
&
& $3.626(1)_{-12.7\%}^{+15.2\%}$
& $+86.3\%$ 
&
& $+80.8\%$\\
\ggNLO{}
& $1.3892(4)_{-13.6\%}^{+15.4\%}$
&
& $3.425(1)_{-12.0\%}^{+13.9\%}$
& $+75.1\%$
&
& $+70.8\%$\\
\hline
& \multicolumn{3}{c|}{$\sigma$\,[fb]} &\multicolumn{3}{c|}{$\sigma/\sigma_{\rm {NLO}}-1$} \\
\hline
\NNLO{}
& $12.88(2)_{-2.2\%}^{+2.8\%}$
&
& $23.55(2)_{-2.6\%}^{+3.0\%}$
& $+14.0\%$
&
& $+18.7\%$\\
\nNNLO{}
& $13.48(2)_{-2.3\%}^{+2.6\%}$
&
& $24.97(2)_{-2.7\%}^{+2.9\%}$
& $+19.3\%$
&
& $+25.8\%$\\
\hline
\end{tabular}
\end{center}
\renewcommand{\baselinestretch}{1.0}
\caption{\label{tableincl} Fiducial cross sections at different perturbative orders and
relative impact on \NLO{} and \ggLO{} predictions, respectively. The quoted uncertainties correspond to scale variations as described in the text, and the numerical integration errors on the previous digit are stated in parentheses; for all ${\rm (n)NNLO}$ results, 
the latter include the uncertainty due the \rcut{} extrapolation~\cite{Grazzini:2017mhc}.}
\end{minipage}
\vspace*{1ex}
\end{table}

\renewcommand{\baselinestretch}{1.0}

We present the fiducial cross sections for $\sqrt{s}=8$ and $13\,\TeV$ at the various perturbative orders in \refta{tableincl}.
In the upper panel the QCD corrections to the quark annihilation channel are reported.
The \NNLO{} corrections to this channel amount to about $+7\%$~($+9\%$) at $\sqrt{s}=8\,(13)\,\TeV$.
In the central panel the loop-induced gluon fusion contribution is shown with its \NLO{} corrections.
Comparing the results with NNLO PDFs used throughout, this contribution 
provides $57\%$~($62\%$) of the full \NNLO{} corrections at $\sqrt{s}=8\,(13)\,\TeV$.
The \NLO{} corrections increase the \ggLO{} result by about $75\%$~($71\%$) at $\sqrt{s}=8\,(13)\,\TeV$.
The contribution of the \qg{} channels is negative, such that the cross section becomes
about $7\%$~($6\%$) larger wrt.\ \ggNLO{} at $\sqrt{s}=8\,(13)\,\TeV$ if contributions  from \qg{} partonic channels are neglected (\ggNLOgg{}).
In the lower panel, the \NNLO{} and \nNNLO{} results are shown.
The impact of the \NLO{} corrections to the loop-induced contribution is to increase the \NNLO{} result by
about $5\%$~($6\%$) at $\sqrt{s}=8\,(13)\,\TeV$.
Corresponding to the above-mentioned numbers, excluding the \qg{} channels would increase the \nNNLO{} prediction by about $1\%$.
The \NNLO{} and \nNNLO{} predictions are marginally compatible within scale uncertainties.

We add a comment on the contribution of diagrams with a Higgs boson: The cuts we are applying essentially select on-shell $Z$ bosons,
thereby forcing the Higgs boson to be off-shell. Nonetheless, our calculation consistently includes also the Higgs diagrams.
The signal--background interference in the $\gg\to ZZ\to 4l$ channel is known to provide a non-negligible contribution~\cite{Caola:2016trd}.
Indeed, we find that with our selection cuts the impact of the Higgs contribution is about $-5\%$ both in the \ggLO{} and \ggNLO{} results.

\begin{figure}[t]
\begin{center}                        
  \begin{tabular}{cc}
    \includegraphics[width=.42\textwidth]{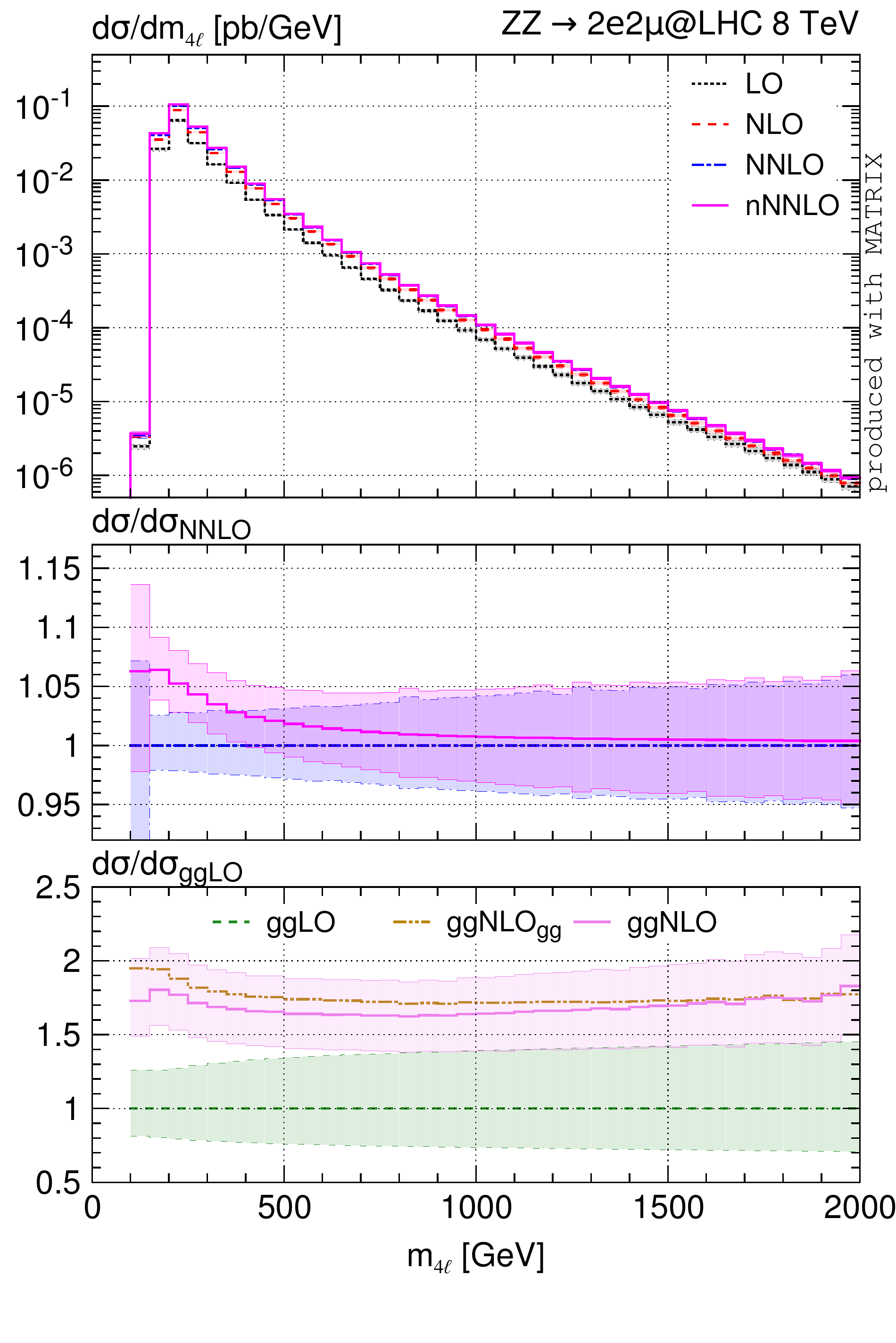} &
    \includegraphics[width=.42\textwidth]{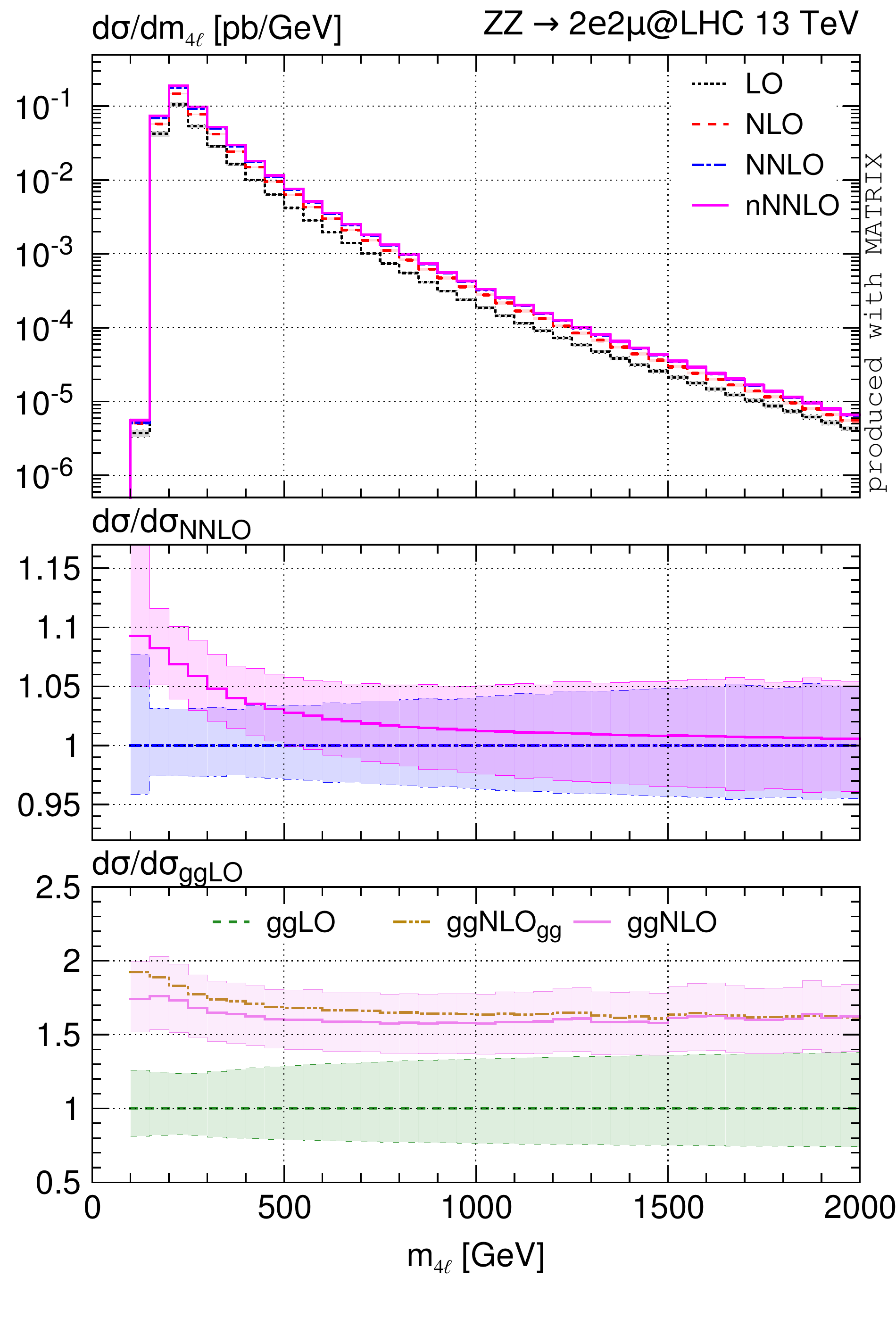}
  \end{tabular}
  \caption{\label{fig:mZZ} Differential distribution in \mfourl{} at $8\,\TeV$ (left) and $13\,\TeV$ (right).}
\end{center}                                                                                                                                                         
\end{figure}      

We now turn to presenting kinematical distributions. Throughout this section, the plots 
are organized according to the following pattern:
There is an upper panel where absolute cross sections at \LO{} (black, dotted), \NLO{} (red, dashed), \NNLO{} (blue, dash-dotted) and \nNNLO{} (magenta, solid) are shown.
In the central panel the \nNNLO{} result with its scale uncertainty is normalised to the central \NNLO{} result.
In the lower panel the \NLO{}/\LO{} $K$-factors of the loop-induced gluon fusion
contribution are shown, with (\ggNLO{}; pink, solid) and without (\ggNLOgg{}; brown, dash-double-dotted) the \qg{} contribution.
The figures on the left show the $8\,\TeV$ results, and the ones on the right the $13\,\TeV$ results.

\begin{figure}                                                                                                                                                                            
\begin{center}                        
  \begin{tabular}{cc}
    \includegraphics[width=.42\textwidth]{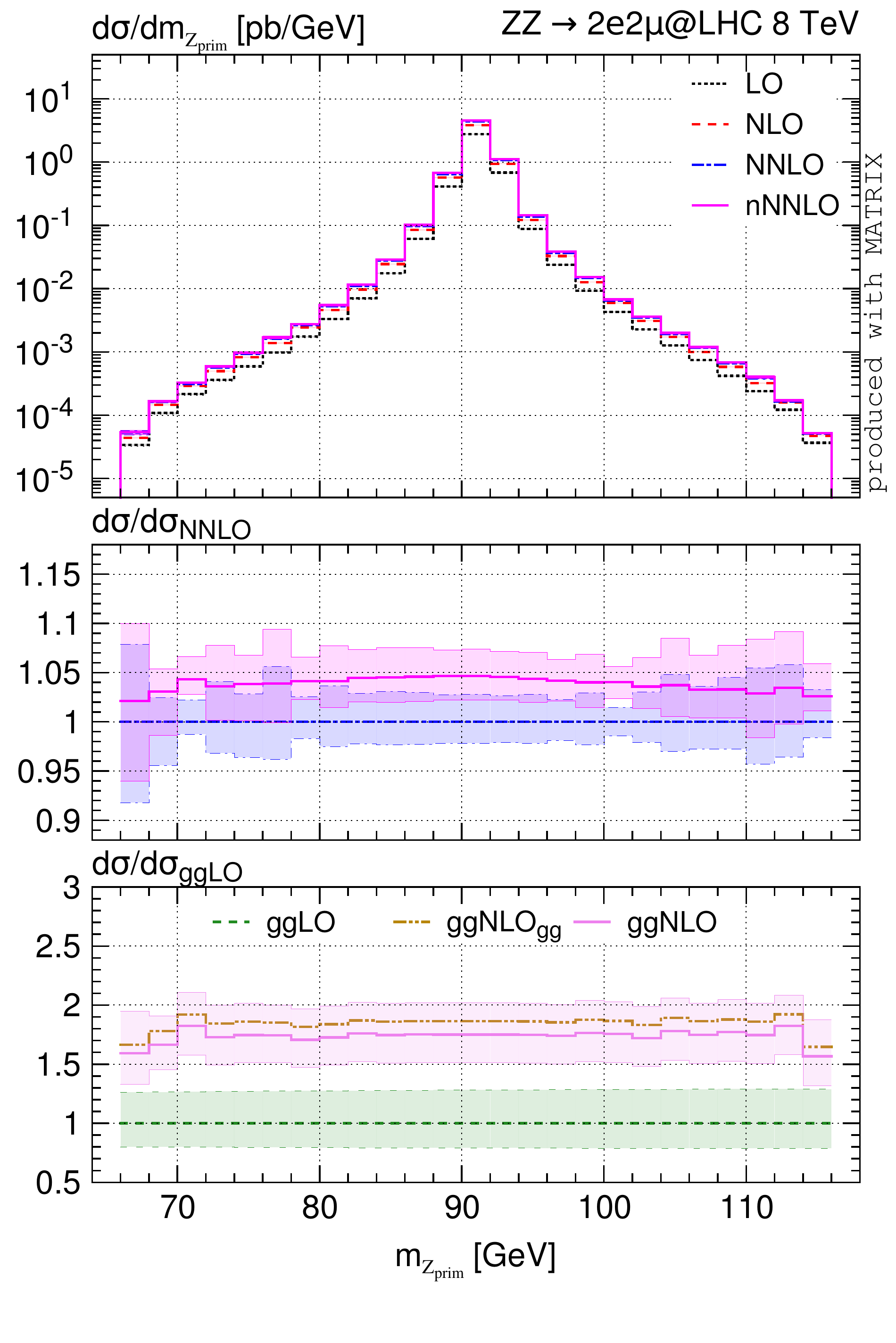} &
    \includegraphics[width=.42\textwidth]{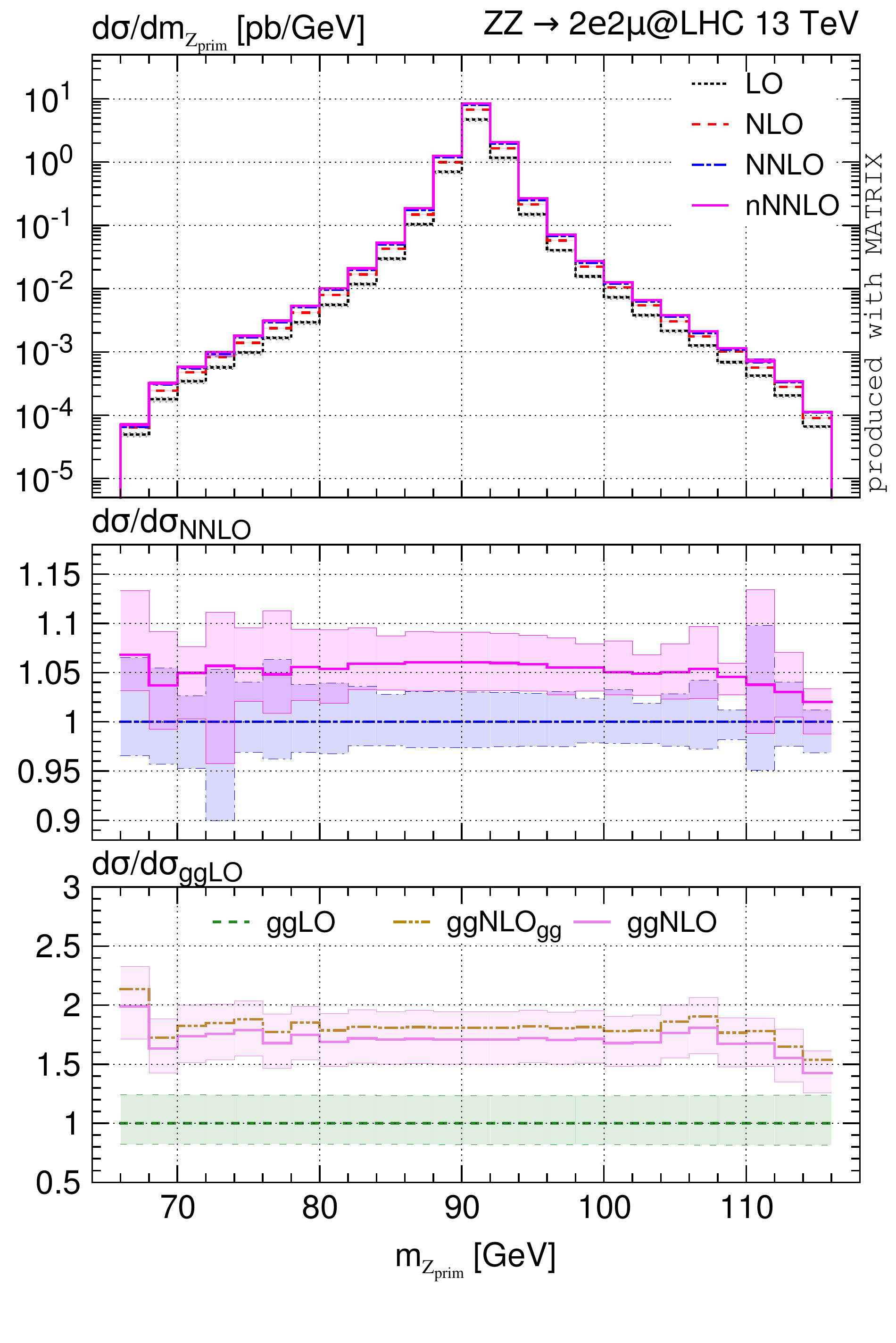}\\
    \includegraphics[width=.42\textwidth]{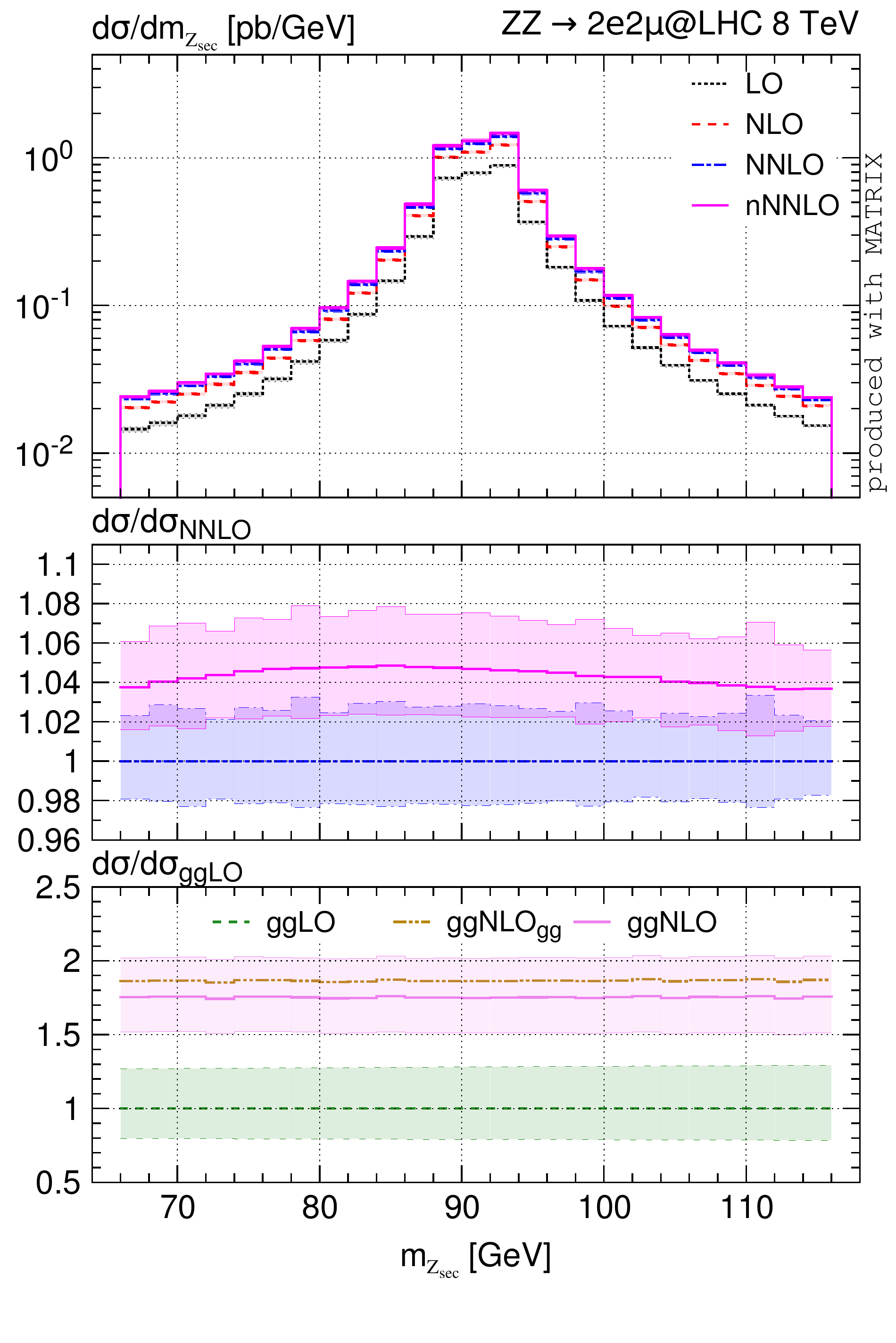} &
    \includegraphics[width=.42\textwidth]{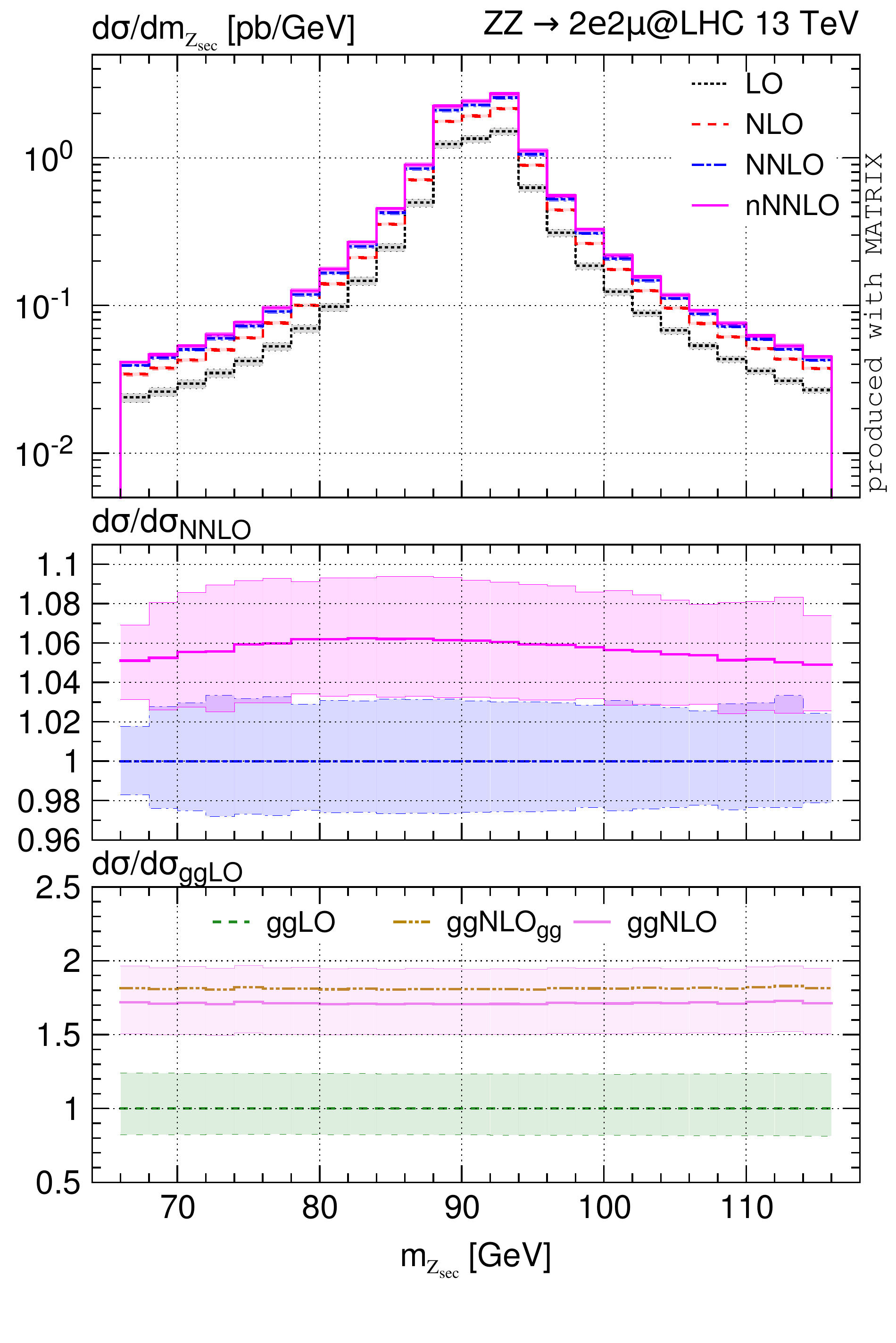}    
  \end{tabular}
  \caption{\label{fig:mZprimsec} Invariant-mass distribution of the OSSF lepton pair closer to (top) and further away from (bottom)
    the $Z$-boson mass at $8\,\TeV$ (left) and $13\,\TeV$ (right).}
\end{center}                                                                                                                                                         
\end{figure}  

We first consider the invariant-mass distribution of the four-lepton system in \fig{fig:mZZ}. 
The impact of the \NLO{} corrections to the loop-induced gluon fusion contribution is largest at small invariant masses:
In the peak region they increase the \NNLO{} cross section by about $5\%$~($7\%$) at $\sqrt{s}=8\,(13)\,\TeV$.
As $\mfourl$ increases, the impact of the \ggNLO{} corrections decreases, and it is
only about $+1\%$ at $\mfourl\sim 1\,\TeV$.
This is not unexpected, since the \gg{} contribution is largest when gluons with smaller $x$ are probed.
On the contrary, the size of the \ggNLO{}/\ggLO{} $K$-factor in the lower panel is relatively stable, with a moderate increase at small $\mfourl$.
In both cases, comparing the \nNNLO{}/\NNLO{} and \ggNLO{}/\ggLO{} ratios, the scale uncertainties do not fully cover the size 
of higher-order corrections in the peak region of the distribution, which demonstrates the importance of the NLO corrections to 
the loop-induced gluon fusion contribution.
The impact of the \qg{} channels on the \ggNLO{}/\ggLO{} $K$-factor is about $-10\%$ at smaller $\mfourl{}$ values, but essentially vanishes in the tail of the $\mfourl{}$ distribution.

In \fig{fig:mZprimsec} we show the invariant-mass distribution of the primary (upper plots) and secondary OSSF lepton pair (lower plots),
ordered by the distance of their invariant masses to the $Z$-boson mass.
Both distributions are limited by the $Z$-mass window cut in the fiducial phase space.
The distribution of the lepton pair
which is less close to $m_Z$ is broader. More precisely, when the invariant mass of the lepton pair is $\mz\pm 20$ GeV, the cross section is suppressed by about four and two orders of magnitude for the primary and secondary lepton pair, respectively. Nonetheless, the impact of QCD corrections is uniform in both cases, and independent of the collider energy. 
The \NNLO{} uncertainty bands barely overlap with the ones of the \nNNLO{} result.

\begin{figure}[t]                                                                                                                                                                             
\begin{center}                        
  \begin{tabular}{cc}
    \includegraphics[width=.42\textwidth]{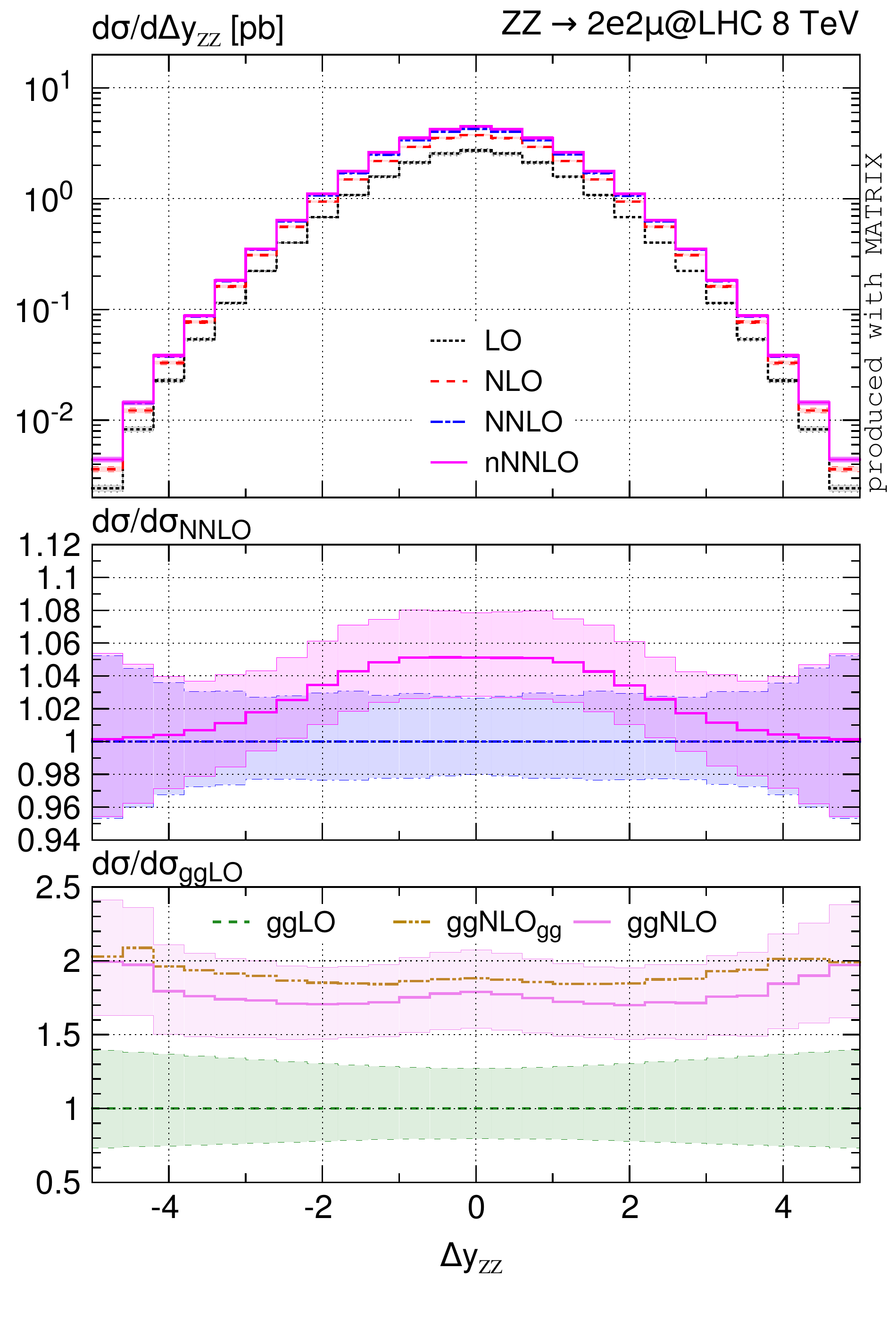} &
    \includegraphics[width=.42\textwidth]{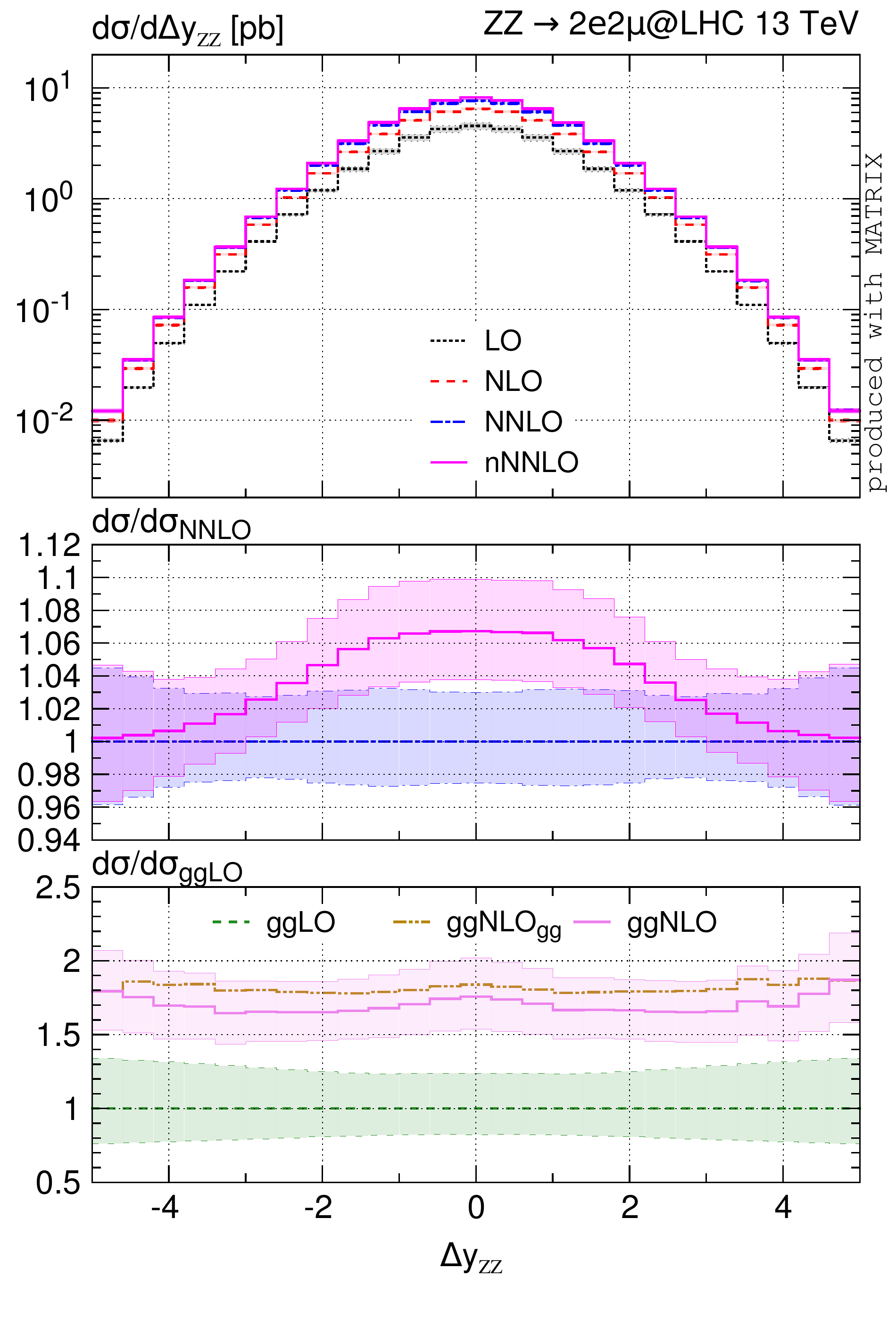}
  \end{tabular}
  \caption{\label{fig:deltayZZ} Differential distribution in $\dyZZ$  at $8\,\TeV$ (left) and $13\,\TeV$ (right).}
  \end{center}                                                                                                                                                         
\end{figure}      

\begin{figure}[t]                                                                                                                                                                             
\begin{center}                        
  \begin{tabular}{cc}
    \includegraphics[width=.42\textwidth]{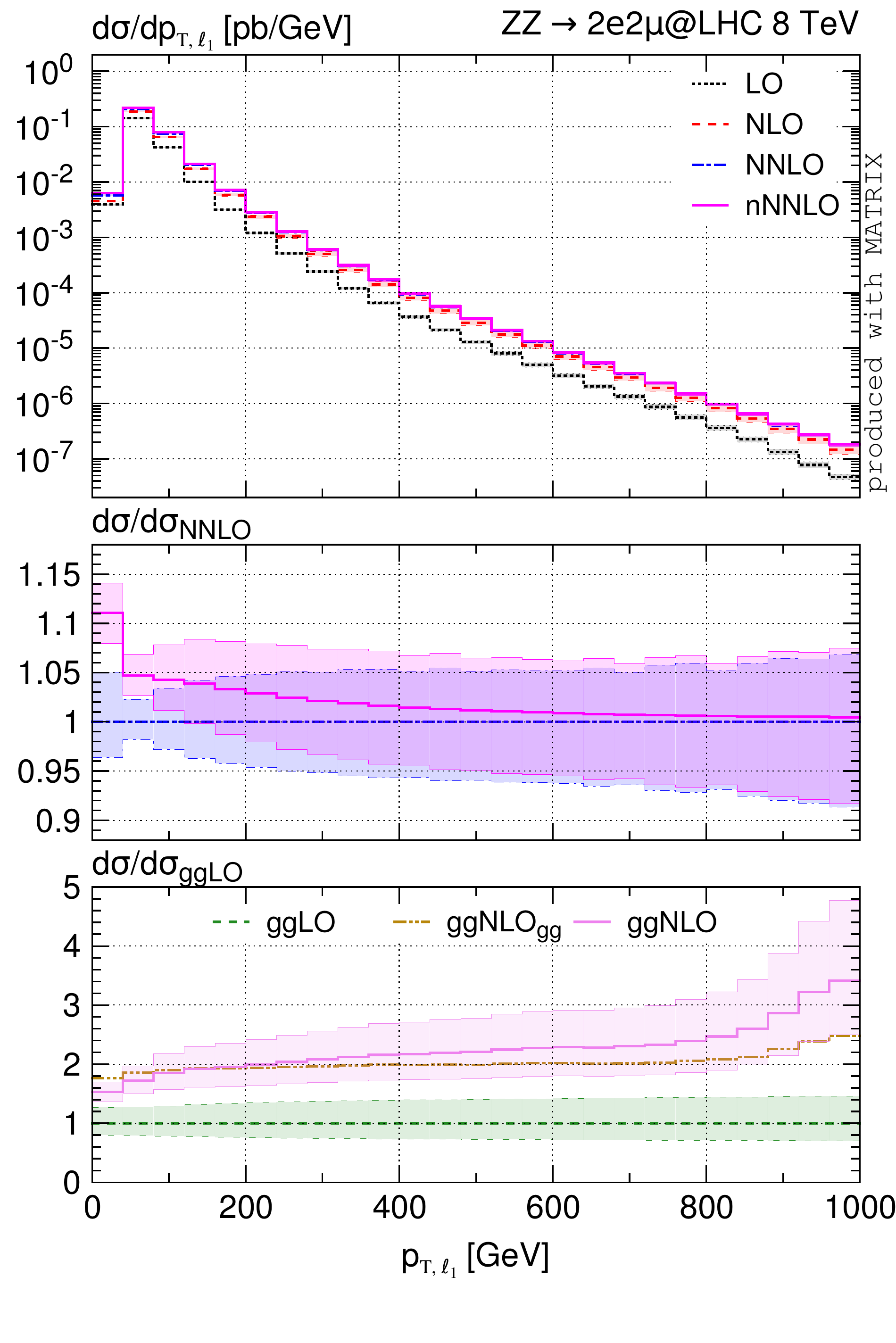} &
    \includegraphics[width=.42\textwidth]{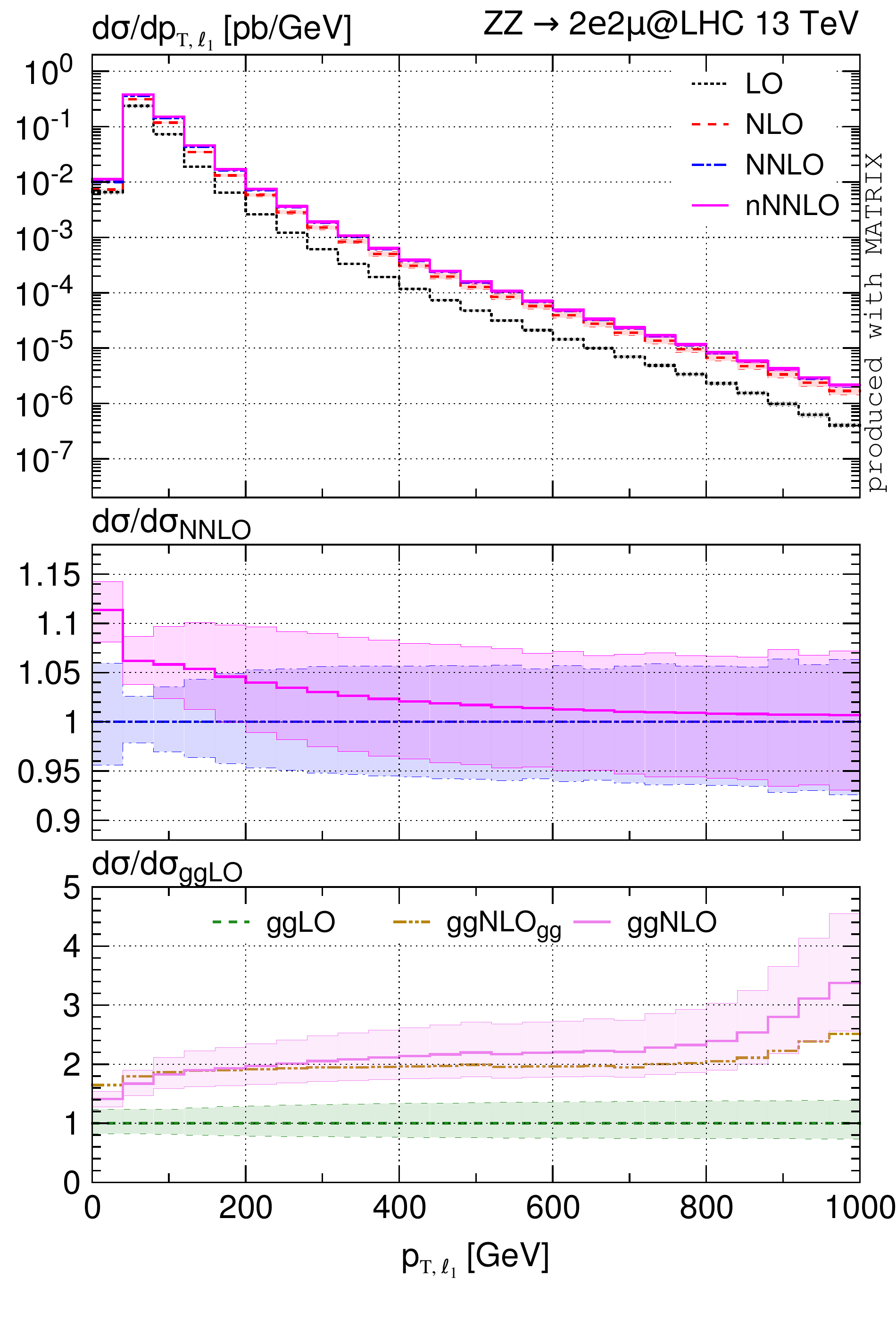}
  \end{tabular}
  \caption{\label{fig:ptl1} Differential distribution in $\ptlone$ at $8\,\TeV$ (left) and $13\,\TeV$ (right).}
\end{center}                                                                                                                                                         
\end{figure} 

\fig{fig:deltayZZ} depicts the distribution in the rapidity separation of the two $Z$ bosons (\dyZZ{}).
The region of small rapidity separations, $|\dyZZ{}|\ltap 1$, is driven by centrally produced $Z$ bosons
and thus relatively small partonic momentum fractions,
which implies that the relative impact of the gluon fusion contribution is most important there.
In this region the impact of the \nNNLO{} corrections is quite uniform and of the order of $+5\%$~($+7\%$) for $\sqrt{s}=8\,(13)\,\TeV$, whereas it successively decreases
in the forward region.
The \ggNLO{}/\ggLO{} $K$-factor is quite flat in \dyZZ{}, and also the relative size of the \qg{} contributions is rather uniform over \dyZZ{}.

In \fig{fig:ptl1} we study the transverse-momentum distribution of the leading lepton~(\ptlone{}). Analogously to \fig{fig:mZZ},
the \nNNLO{} corrections are maximal at small \ptlone{},
and they decrease with the value of \ptlone{}: They are about $+5\%$ in the peak region and drop to about $+1\%$ at \mbox{$\ptlone\sim 500\,{\rm GeV}$}.
Also here the perturbative uncertainties in the first bins do not cover the difference between 
the \NNLO{} and \nNNLO{} predictions.
In contrast to \fig{fig:mZZ}, the \ggNLO{}/\ggLO{} $K$-factor of the gluon fusion contribution becomes larger with the value of \ptlone{},
from about 1.7 in the peak region to about 2.2 at $\ptlone\sim 500$ GeV, and it further increases for larger \ptlone{} values.
It is interesting to notice that the impact of the newly included \qg{} channels 
on the \ggNLO{} corrections depends on the value
of \ptlone{}, with roughly $-10\%$ in the peak region and $+10\%$ 
for $\ptlone\sim 500$ GeV, thereby affecting the shape of the distribution.

\begin{figure}                                                                                                                                                                           
\begin{center}                        
  \begin{tabular}{cc}
    \includegraphics[width=.42\textwidth]{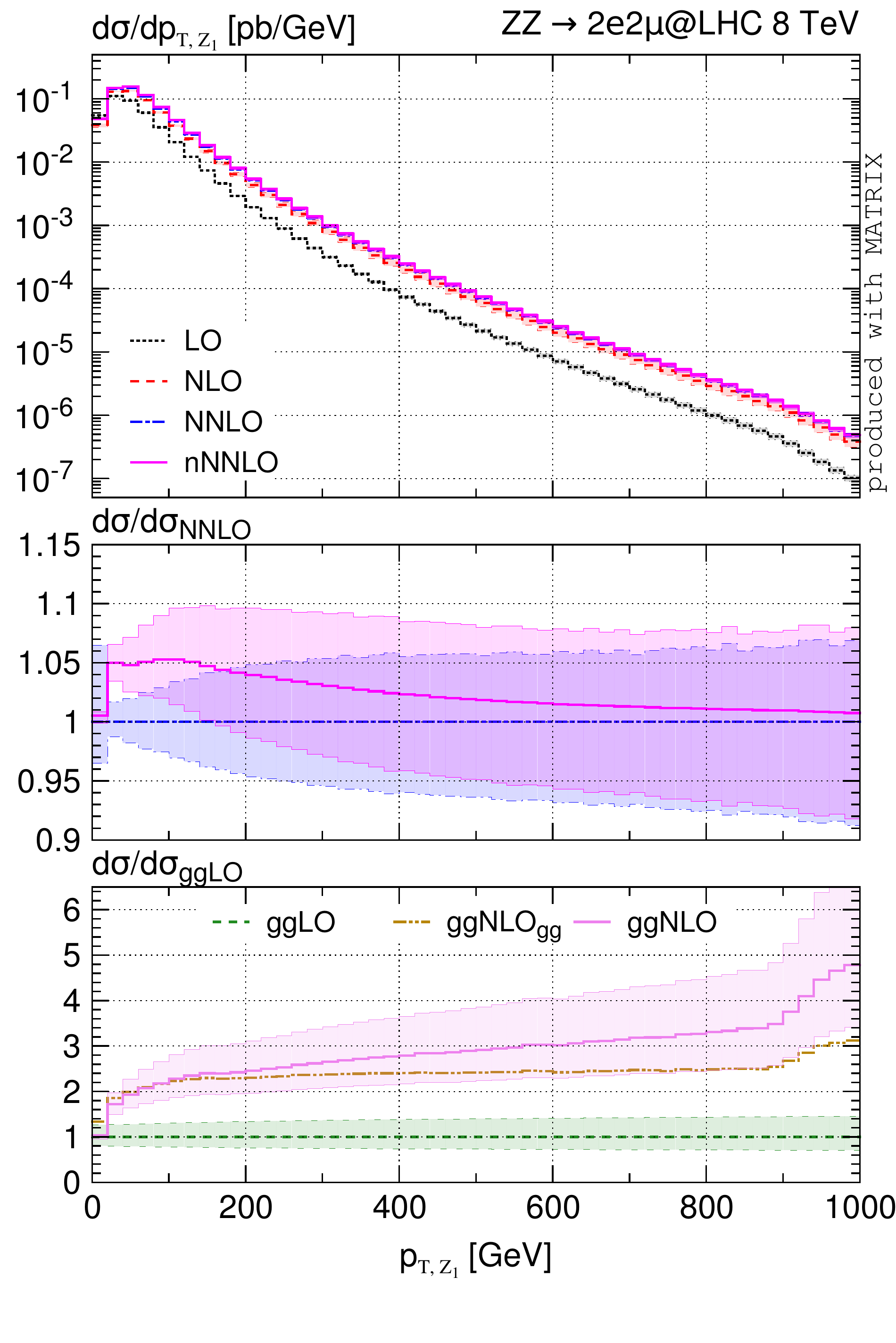} &
    \includegraphics[width=.42\textwidth]{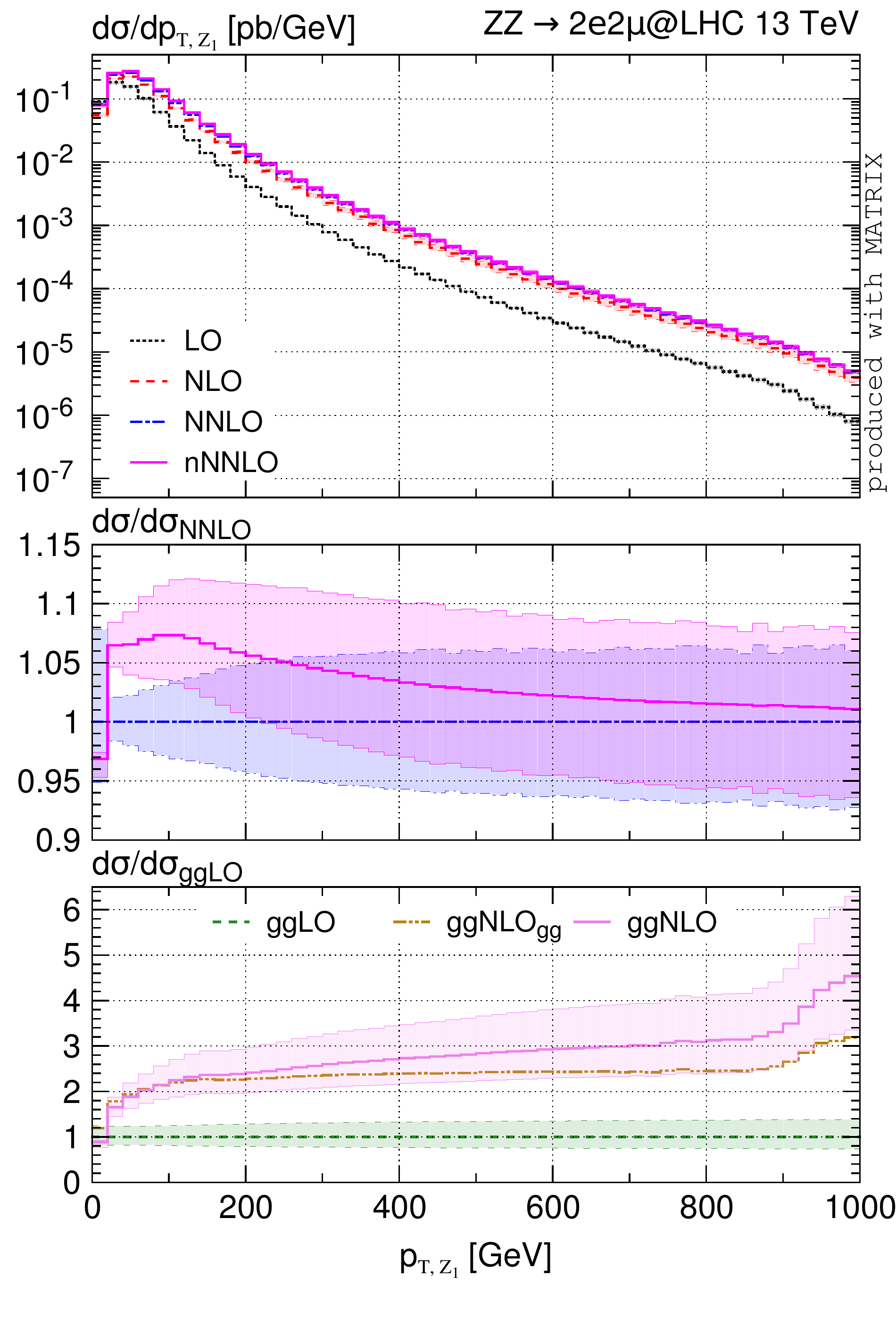}\\
    \includegraphics[width=.42\textwidth]{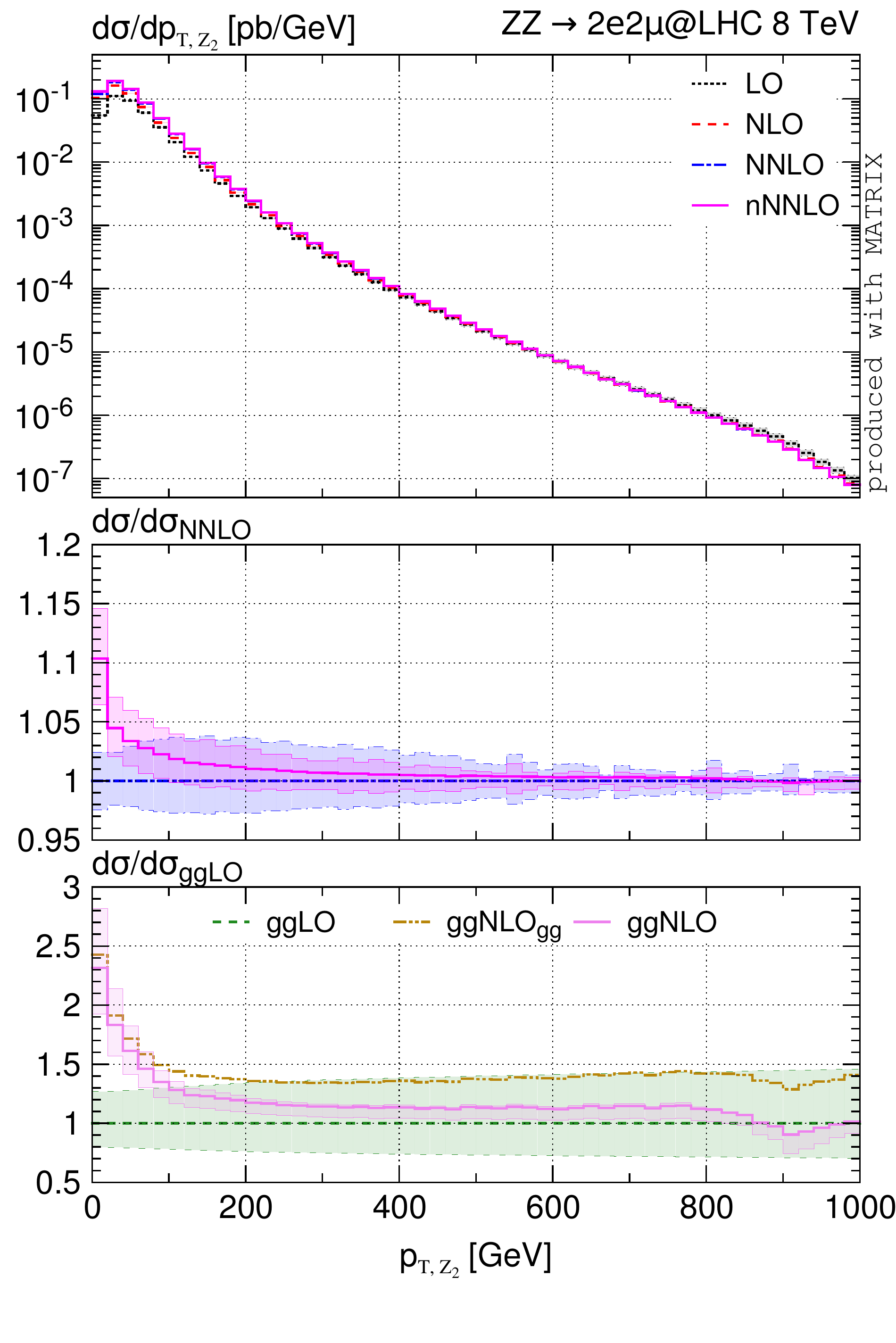} &
    \includegraphics[width=.42\textwidth]{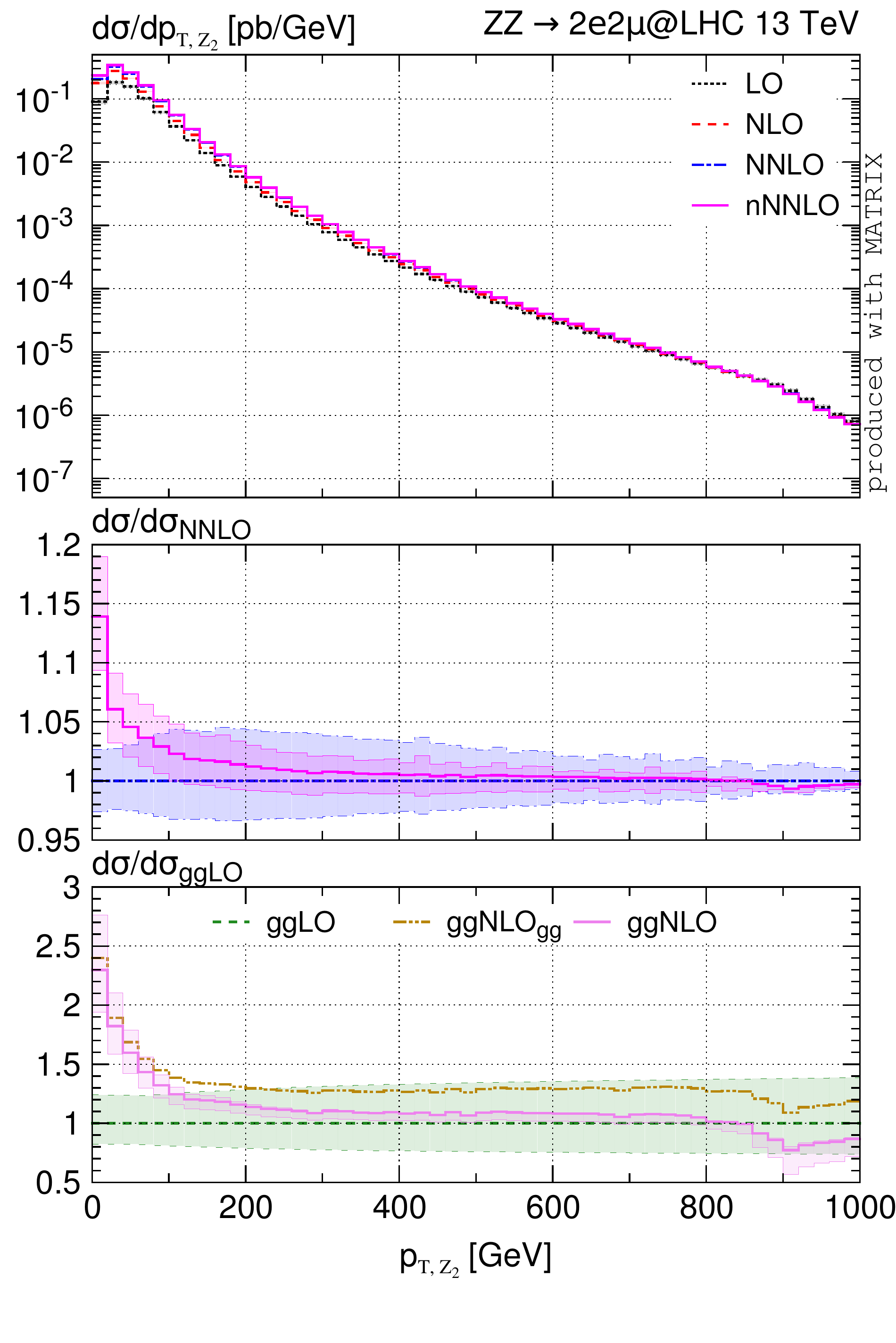}
  \end{tabular}
  \caption{\label{fig:ptZ} Transverse-momentum spectra of the leading (top) and subleading (bottom) $Z$ boson at $8\,\TeV$ (left) and $13\,\TeV$ (right).}
\end{center}                                                                                                                                                         
\end{figure} 

In \fig{fig:ptZ} the transverse-momentum distributions of the leading (\ptzone{}) and subleading (\ptztwo{}) reconstructed $Z$ bosons
are shown.
The most prominent feature we notice are the large \NLO{}/\LO{}
corrections to the loop-induced gluon fusion channel in case of the harder $Z$ boson,
which significantly increases with \ptzone{}: The enhancement
is by a factor of about 2.8 at $p_T=500$ GeV.
This is due to the fact that the phase-space region with one hard $Z$ boson
is dominantly populated by events with a jet recoiling against this $Z$ boson, while the
other $Z$ boson is relatively soft. 
Such large corrections are absent in the \ptztwo{} distribution: The phase-space region where
both $Z$ bosons have large transverse momenta is naturally dominated
by topologies with the two $Z$ bosons recoiling against each other, which are already present at \LO{}
and thus do not give rise to exceptionally large corrections.
This situation also explains the opposite behaviour of the \NLO{}/\LO{} $K$-factor in \ptztwo{} which 
continuously decreases with \ptztwo{}.
The previous statements are not specific to the loop-induced gluon fusion channel: We observe the same 
features also for the NLO corrections to the quark annihilation channel.

Also for the transverse-momentum distributions of the $Z$ bosons
the importance of the \qg{} channels in the \ggNLO{} 
result is evident: The \ptzone{} shape is clearly modified due to a negative \qg{} 
contribution at small \ptzone{}, and a positive \qg{} contribution in the tail of the distribution. At large
\ptztwo{} the contribution of the \qg{} channels is as large as the one of the \gg{} channel.
However, they have opposite signs such that they compensate each other and the \ggNLO{} corrections almost vanish, 
whereas, neglecting \qg{} contributions, the \ggNLOgg{} corrections show an increase of roughly $40\%$ wrt.\ \ggLO{} instead.
\NNLO{} scale uncertainties 
at small \ptzone{} and \ptztwo{} typically do not cover the sizeable \nNNLO{} corrections.

Another eye-catching feature we observe in \fig{fig:ptZ} is the significant drop of the transverse-momentum distribution
of both the leading and subleading $Z$ boson above $\ptzi\sim 900$ GeV ($i\in\{1,2\}$).
This is due to the interplay between the large transverse momentum of the parent $Z$ boson, which makes the corresponding lepton pair boosted,
and the $\dRll>\dRll^{\rm min}$ cut in the fiducial phase space ($\ell\in\{e,\mu\}$, $\dRll^{\rm min}=0.2$).
Indeed, if the transverse momentum of the parent $Z$ boson fulfills the condition
\begin{equation}
  \ptzi\gtap \frac{\sqrt{2}\,\mz}{\sqrt{1-\cos\dRll^{\rm min}}}\sim 900~{\rm GeV}\,,
  \end{equation}
the lepton pair is forced to be produced off-shell, and
as a consequence the cross section is strongly suppressed.
Note that this effect is independent of the collider energy.

\section{Summary}
\label{sec:summary}

We have calculated the \nlo{} \qcd{} corrections to the loop-induced gluon fusion
contribution for \zz{} production in the four-lepton channel. 
Our predictions include, for the first time, also (anti)quark--gluon partonic channels.
We have combined these results with state-of-the-art \nnlo{} \qcd{} corrections to the quark annihilation channel,
yielding an approximation of the full N$^3$LO \qcd{} corrections for
\zz{} production, denoted by \nNNLO{}.

We have performed an extensive validation against existing results for the \nlo{} 
cross section of the loop-induced gluon fusion contribution in the literature~\cite{Caola:2015psa,Alioli:2016xab}.
Overall, we find decent agreement of the total rates and distributions 
by adopting the respective setups, thereby neglecting contributions from \qg{} channels and massive top-quark loops.

We have presented a comprehensive study of the \nNNLO{} corrections, 
the size of the \ggNLO{}/\ggLO{} $K$-factors of the loop-induced gluon fusion contribution, and 
the impact of the newly computed \qg{} channels. The main conclusions can be summarized
as follows:
\begin{itemize}
\item The loop-induced gluon fusion contribution is sizeable. It makes up roughly $57\%$~($62\%$) of
the full $\mathcal{O}(\as^2)$ corrections to \zz{} production, and yields roughly $6\%$~($9\%$) of the 
\NNLO{} cross section at $8\,(13)\,\TeV$ collider energy. 
Hence, its \NLO{} QCD corrections are important, with $K$-factors of $\gtrsim 1.7$ wrt.\ \ggLO{}: The \nNNLO{} cross section is about
$5\%$ ($6\%$) larger than the \NNLO{} one at $\sqrt{s}=8\,(13)\,\TeV$, and their perturbative 
uncertainties barely overlap.
\item The \qg{} channels have a negative effect of about $10\%$ on the \ggNLO{}/\ggLO{} $K$-factor 
of the loop-induced gluon fusion contribution, which yields roughly a $1\%$ decrease of
the \nNNLO{} cross section.
\item The \nNNLO{} corrections can have a non-trivial impact on differential
distributions. Due to the nature of the loop-induced gluon fusion production mechanism, 
and the dominance of the gluon densities at small $x$, 
the \nNNLO{} corrections provide sizeable effects in dominant phase-space 
regions of the observables we investigated:
at small invariant-masses of the four lepton system, in 
the central region of the rapidity difference between the two reconstructed $Z$ bosons, 
and at small transverse momenta of the leptons and $Z$ bosons.
\item In these regions, where the computed corrections are largest, \nNNLO{} and \NNLO{} scale-uncertainty bands barely overlap,
which demonstrates the importance 
of including the \NLO{} QCD corrections to the loop-induced gluon fusion contribution.
\item For the transverse-momentum distributions of the leptons and the reconstructed $Z$ bosons
the newly computed contributions of the \qg{} channels have a significant impact on the shapes
of the \ggNLO{} spectra.
\end{itemize}

The \nlo{} \qcd{} corrections to the loop-induced gluon fusion contribution have been 
implemented in our parton-level Monte Carlo code \Matrix{} to provide \nNNLO{} 
cross sections for charge-neutral diboson production processes. 
The consistent combination of state-of-the-art predictions for quark annihilation and loop-induced gluon fusion
production mechanisms within a single tool may turn out to be particularly useful for the 
experimental analyses. \Matrix{} can be used not only to obtain the best QCD prediction for diboson 
cross sections, but also to estimate perturbative uncertainties
consistently from simultaneous scale variations within the two contributions. 
The inclusion of \NLO{} EW corrections in these calculations to obtain ultimate perturbative 
accuracy in diboson predictions is left for future work.

\noindent {\bf Acknowledgements}

We thank Simone Alioli, Fabrizio Caola, and Lorenzo Tancredi for helpful discussions, and for providing 
details on their computations and results. We are particularly indebted to Jonas Lindert
for making private \OpenLoops{} amplitudes for the loop-induced channel available to us,
notably spin and colour correlators and several approximations with and without top-quark contributions,
and Simone Alioli for private communication
of the statistical uncertainties to the results of \citere{Alioli:2016xab}.
Moreover, we want to thank Jean-Nicolas Lang for his help on interfacing \Recola{} and validating
the \OpenLoops{} results.
This work is supported in part
by the Swiss National Science Foundation (SNF) under
contracts CRSII2-141847 and 200020-169041.
The work of MW is supported by the ERC Consolidator Grant 614577 HICCUP,
and that of SK by the ERC Starting Grant 714788 REINVENT.
\renewcommand{\em}{}
\bibliographystyle{apsrev4-1}
\bibliography{zznlogg}

\begin{thebibliography}{79}
\bibitem [{mun()}]{munich}%
  \BibitemOpen
  \href@noop {} {}\bibinfo {note} {{\sc Munich} is the abbreviation of
  ``MUlti-chaNnel Integrator at Swiss (CH) precision''---an automated parton
  level NLO generator by S.~Kallweit. In preparation.}\BibitemShut {Stop}%
\end{thebibliography}
\end{document}